\documentstyle[12pt,epsfig]{article}
\title{{\LARGE\bf Problems and stoppers for $\gamma\gamma,
    \gamma\mu, \mu p$ colliders using very high energy muons.}
  \thanks{Invited talk at the Workshop Studies on Colliders and
    Collider Physics at the Highest Energies: Muon Colliders at 10 TeV
    to 100 TeV, 27 September - 1 October, 1999 Montauk, New York, USA,
    be published by the American Institute of Physics.  }  }

\author{Valery Telnov, \\
  {\small\it Budker Institute of Nuclear Physics, 630090 Novosibirsk, 
  Russia}\thanks{email:telnov@inp.nsk.su}} 

\begin{document}
\newcommand{\EP}{\mbox{e$^+$}}
\newcommand{\EM}{\mbox{e$^-$}}
\newcommand{\EPEM}{\mbox{e$^+$e$^-$}}
\newcommand{\EMEM}{\mbox{e$^-$e$^-$}}
\newcommand{\MP}{\mbox{$\mu$p}}
\newcommand{\EE}{\mbox{ee}}
\newcommand{\GG}{\mbox{$\gamma\gamma$}}
\newcommand{\GE}{\mbox{$\gamma$e}}
\newcommand{\GM}{\mbox{$\gamma\mu$}}
\newcommand{\TEV}{\mbox{TeV}}
\newcommand{\GEV}{\mbox{GeV}}
\newcommand{\LGG}{\mbox{$L_{\gamma\gamma}$}}
\newcommand{\LMM}{\mbox{$L_{\mu\mu}$}}
\newcommand{\EV}{\mbox{eV}}
\newcommand{\CM}{\mbox{cm}}
\newcommand{\MM}{\mbox{mm}}
\newcommand{\NM}{\mbox{nm}}
\newcommand{\MKM}{\mbox{$\mu$m}}
\newcommand{\SEC}{\mbox{s}}
\newcommand{\CMS}{\mbox{cm$^{-2}$s$^{-1}$}}
\newcommand{\MRAD}{\mbox{mrad}}
\newcommand{\IND}{\hspace*{\parindent}}
\newcommand{\E}{\mbox{$\epsilon$}}
\newcommand{\EN}{\mbox{$\epsilon_n$}}
\newcommand{\EI}{\mbox{$\epsilon_i$}}
\newcommand{\ENI}{\mbox{$\epsilon_{ni}$}}
\newcommand{\ENX}{\mbox{$\epsilon_{nx}$}}
\newcommand{\ENY}{\mbox{$\epsilon_{ny}$}}
\newcommand{\EX}{\mbox{$\epsilon_x$}}
\newcommand{\EY}{\mbox{$\epsilon_y$}}
\newcommand{\BI}{\mbox{$\beta_i$}}
\newcommand{\BX}{\mbox{$\beta_x$}}
\newcommand{\BY}{\mbox{$\beta_y$}}
\newcommand{\SX}{\mbox{$\sigma_x$}}
\newcommand{\SY}{\mbox{$\sigma_y$}}
\newcommand{\SZ}{\mbox{$\sigma_z$}}
\newcommand{\SI}{\mbox{$\sigma_i$}}
\newcommand{\SIP}{\mbox{$\sigma_i^{\prime}$}}
\date{}
\maketitle
\begin{abstract}
   
   It is well known that at linear \EPEM\ (\EE) colliders using laser
backscattering one can obtain colliding \GG, \GE\ beams with 
energy and luminosity comparable to those in \EPEM\ collisions. In
this paper, it is explained why this can not be done at high energy
muon colliders. Due to several physics reasons the \GG\ luminosity
is suppressed here by a factor of $10^{14}$ !  Another option --
$\gamma$'s from a linear collider and muons from a muon collider -- is
also discussed (and has no sense either). Of course, one can study $\gamma^*
\mu$ and $\gamma^*\gamma^*$ interactions at muon colliders in
collisions  with virtual photons as it is done now at \EPEM\
storage rings. Muon-proton colliders are attractive only if the proton beam
is cooled and has the same parameters as the muon beam, in which case  $L_{\mu
p} \sim \LMM$.

\end{abstract}

\section{Introduction}
     
     Firstly, I would like to explain the origin of this talk.  Two
weeks ago the chairman of our workshop Bruce King have sent me e:mail with the
request to give a plenary talk on ``prospects for very high energy \GG\
or $\gamma\mu$ colliders driven by the muon beams'', he added that
``even if it is impractical it would still be nice if you could give a
brief explanation''.

  I have agreed to give such a talk but only without the word
  ``prospects'' in the title because I do not see  any prospects here, only
  stoppers. Nevertheless, this physics  is very interesting, and it is
  pleasure to me to tell briefly about high energy photon colliders based
  on \EPEM\ (ee) linear colliders and explain why such photon colliders are
  completely impractical with muons.

The third combination of colliding particles, $\mu$p, is also discussed
here very briefly.

\section{Photon Colliders based on  linear \EE\ colliders}

 As you know, to explore the energy region beyond LEP-II, linear
\EPEM\ colliders (LC) in the range from a few hundred GeV to about 1.5
TeV and higher are under intense study around the
world~\cite{NLC,TESLA,JLC,CLIC}.

 Beside \EPEM\ collisions, linear colliders provide a unique
possibility for obtaining \GG, \GE\ colliding beams with  energies and
luminosities comparable to those in \EPEM\
collisions~\cite{GKST81,GKST83,GKST84,TEL90,TEL95}.  High energy
photons for these collisions can be obtained using Compton scattering
of laser light on high energy electrons. This idea is based on the
following facts:

\begin{itemize}
\item Unlike the situation in storage rings, in linear colliders each beam
is used only once.
\item Using an optical laser with reasonable parameters (flash energy
  of 1 to 5 J) one can ``convert'' almost all electrons to high energy
  photons;
\item The energy of scattered photons is close into the energy of
  initial electrons.
\end{itemize}

Each one of these items is vital for obtaining \GG, \GE\ collisions
at  energies and luminosities comparable to those in parental
electron-electron collisions.~\footnote{Here we do not discuss ``photon
colliders'' based on collisions of virtual photons. This possibility
always exists, however the luminositis and energies are
considerably smaller than those in parental ee  collisions, see sect.\ref{sum}}

The physics at high energy \GG,\GE\ colliders is very rich and no less
interesting than with pp or \EPEM\ collisions. This option has been
included in the pre-conceptual design reports of  LC
projects~\cite{NLC,TESLA,JLC}, and work on  full conceptual designs
is under way.  Reports on the present status of photon colliders can be found
elsewhere~\cite{Tfrei,Tsit1}.

   Well, can we make similar photon colliders on the  basis of muon colliders?
What is the difference? 

\section{\GG,\GM\ colliders based on high energy $\mu\mu$ colliders}
\subsection{Multi-pass collisions}

  At muon colliders two bunches are collided about 1000 times, which
is one of the advantages over linear \EPEM\ colliders where beams are
collided only once.  However, if one tries to convert muons into high
energy photons (by whatever means), the resulting \GG\ luminosity will
be smaller than that in $\mu\mu$ collisions at least by a factor of
1000.  This argument alone  sufficient to give up the idea
of \GG\ colliders based on high energy muon colliders. However, at
this workshop F.Zimmermann proposed the idea of one pass muon
colliders. So, I will continue enumeration of stoppers.

\subsection{Laser wave length}

 The required wave length follows from the kinematics of Compton
scattering~\cite{GKST83}. In the conversion region a laser photon with
the energy $\omega_0$ scatters at a small collision angle (head-on)
on a high energy electron(muon) with the energy $ E_0$.  The maximum
energy of scattered photons (in direction of electrons) is given by
\begin{equation}
\omega_m=\frac{x}{x+1}E_0; \;\;\;\;
x=\frac{4E_0\omega_0}{m^2c^4},
\end{equation}
where $m$ is the mass of the charged particle. In order to obtain 
photons with the energy comparable to that of initial particles, say, 80 \%,
one needs $x\sim 4$, or the energy of laser photons
\begin{equation}
\omega_0 \sim m^2c^4/E_0.
\label{wave}
\end{equation}
 The corresponding laser wave length is then 
\begin{equation} 
\lambda \sim 5E_0[\TEV]\; \MKM\ \;\;\; \mbox{for electron beams;}
\end{equation}
\begin{equation}
\lambda \sim 0.12E_0[\TEV]\; \NM\  \;\;\; \mbox{for muon beams}. 
\end{equation}
 So, one can use optical lasers to make high energy photons by means
of backward Compton scattering on electron beams, while at muon
colliders one would have to use X-ray lasers!

\subsection{Flash energy}
  
  The probability of Compton scattering for an beam particle in the laser
target $p \sim n\sigma_C l$, where $n, l,\sigma_C$ are the density of
the laser target, its length and the Compton cross section,
respectively. The density $n \sim (A/lS)/\omega_0$, where $A$ is the
laser flash energy and $S$ is the cross section of the laser beam
which should be larger than that of the muon beam.~\footnote{In the
case of the electron LC, where optical photons are used, the laser
spot size is determined by diffraction: $a_{\gamma}\sim \sqrt{\lambda
l/4\pi}$ which is several \MKM\ for LC electron beams~\cite{TEL90}. At
muon colliders, the required wave length is much shorter and
diffraction can be neglected.}  The Compton cross section for muon at
$x=4$ is about \cite{GKST83}
\begin{equation}
\sigma_C(x=4)   \sim \pi r_e^2\left(\frac{m_e}{m_{\mu}}\right)^2,   
\end{equation}
where $r_e=e^2/m_e c^2$ is the classical radius of the electron.

  From the above relations we get the required laser flash energy (for
  $p\sim 1$)
\begin{equation}
A \sim (S/\sigma_C)\omega_0 = \frac{S}{\pi r_e^2
E_0}\left(\frac{m_{\mu}}{m_e}\right)^4 m_e^2c^4=
1.5\times10^{-3}\frac{S[\MKM^2]}{E_0[\TEV]}\left(\frac{m_{\mu}}{m_e}\right)^4
\;\; \mbox{Joule}.
\end{equation}
At the muon collider with $E_0=50$ TeV, $S=1$ $\MKM^2$ one needs the
X-ray laser with the flash energy $10^5$ J and the wave length of 6
nm (see eq. \ref{wave}). This is certainly impossible.  Beside this
``technical'' problem, there are even more fundamental stoppers
for photon colliders based on muon beams, see below.

\subsection{\EPEM\ pair creation in the conversion region}

  Beside the Compton scattering at the conversion region, at muon
  colliders there is another competing process: \EPEM\ pair creation in
  collision of laser photons with the high energy  muons, $\gamma \mu \to \mu
  \EPEM$. The ratio of the cross sections
\begin{equation}
\frac{\sigma_{\gamma \mu \to \mu \EPEM}}{\sigma_{\gamma \mu \to \gamma
\mu}} \sim \frac{\frac{28\alpha
r_e^2}{9}\ln{\frac{4E_0\omega_0}{m_{e}m_{\mu}c^4}}} {\pi
r_e^2(m_e/m_{\mu})^2} \sim
7\times10^{-3}\left(\frac{m_{\mu}}{m_e}\right)^2
\ln{\left(\frac{m_{\mu}}{m_e}x\right)} \sim 2000\;\; \mbox{at} \;\; x = 4.
\end{equation}
So, high energy photons are produced with a very small probability,
less than 1/1000 !  In all other cases muons lose their energy via
creation of \EPEM\ pairs.  This effect alone suppresses the attainable
\GG\ luminosity at muon colliders by a factor of more than $10^6$ !

\subsection{Coherent pair creation}

   OK, the yield of high energy photons from the conversion region is
   very small, but this is not the whole story. What happens to the
   ``happy'' photons at the interaction region?  They will be
   ``killed'' by the process of coherent \EPEM\ pair creation in the
   field of the opposing muon beam. This process restricts the
   luminosity of photon colliders based on electron linear
   colliders~\cite{TEL90,TEL95,TSB2}.\footnote{For an LC with the energy
   below about 1 TeV this effects is still not very important and
   one can obtain, in principle, $\LGG\ > L_{\EPEM}$} The effective
   threshold of this process $\Upsilon =
   \frac{\omega}{m_ec^2}\frac{B}{B_0} \sim 1$, where $\omega$ is the
   photon energy, $B$ is the beam field, $B_{0} = \alpha e/r_e^2 \sim
   4.4\times 10^{13}$ Gauss.  For the ``evolutionary'' $2E=100$ TeV
   muon collider (see the B.King's tables) with $N=0.8\times10^{12}$,
   $\sigma_z=2.5$ mm, $\sigma_{x,y}=0.2$ \MKM\ and a photon energy
   40 TeV we have $\Upsilon \sim 180$. Using formulae given in 
   ref.\cite{TEL90}, one can find the probability of \EPEM\ pair
   creation during the  bunch collision: it is very high, about
   200. This means that only about 1\% of high energy photons will
   survive in  beam collisions and contribute to the \GG\
   luminosity.

\subsection{Summary on  \GG, \GM\  colliders based on high energy muons 
\label{sum}}

1) The laser required for conversion of 50 TeV muons into high energy
   photons should have flash energy $A \sim 10^5$ J and wave length
   $\lambda ~\sim 5$ nm.  This is impossible.

2) The achievable \GG\ luminosity
\begin{equation}
\LGG/\LMM\ \sim \frac{1}{1000}\times \left(\frac{1}{2000}\right)^2 \times
\left(\frac{1}{100}\right)^2 = 2.5\times 10^{-14}\;!
\end{equation}
Here the first factor is due to the one pass nature of photon
colliders, the second one is due to the dominance of \EPEM\ creation at
the conversion region (instead of Compton scattering), and the third
one is due to  coherent pair creation at the interaction
region. All clear. One can forget about \GG\ (and \GM\ too) colliders
based on  high energy muon beams.

  However, \GG,\GM\ interactions can be studied at muon colliders in
collisions of virtual photons (without $\mu \to \gamma$
conversion). The luminosities in such collisions are~\cite{TEL90}
\begin{equation}
L_{{\gamma}^*{\gamma}^*} \sim 10^{-2}\LMM\  \;\;\;\;\;W_{\GG} > 0.1\times2E_0 
\end{equation}
\begin{equation}
L_{{\gamma}^*{\gamma}^*} \sim 10^{-4}\LMM\  \;\;\;\;\;W_{\GG} > 0.5\times2E_0. 
\end{equation}
\begin{equation}
L_{{\gamma}^*\mu} \sim 0.15\LMM\  \;\;\;\;\;W_{\gamma \mu} > 0.1\times2E_0 
\end{equation}
\begin{equation}
L_{{\gamma}^*\mu} \sim 0.05\LMM\  \;\;\;\;\;W_{\gamma \mu} > 0.5\times2E_0. 
\end{equation}

\section{\GM\  collisions at LC--muon colliders}

One can also consider \GM\ colliders where high energy photons are
produced at LC (on electrons) and then are collided with high energy
muon beams. This option also has no sense for  several reasons: \\
a) $N_e \sim 10^{-2}N_{\mu}$; \\ 
b) loss of photons at the IP due to
coherent \EPEM\ pair creation; \\ 
c) none of the LCs have the pulse structure
of muon colliders (almost uniform in time), which results in a factor 100
times loss in  luminosity. \\ 

All factors combined give $L_{\gamma \mu} < 10^{-5}\LMM$.  Such $\gamma\mu$
collider has no sense; besides, $\gamma\mu$ collisions can be studied for
free with much larger luminosities in $\gamma^*\mu$ collisions (see the
end of the previous section).

\section{ \MP\ colliders}

   Let us first consider collisions of the LHC proton beams with muon
beams of a 100 TeV muon collider.  Without special measures, the
luminosity in such collisions is lower than that in pp collisions at
LHC due to larger distance between bunches at muon colliders (smaller
collision rate).

\begin{equation}
L_{\mu p} \sim L_{pp} \times (\nu_{\mu}/\nu_p) \sim 10^{-3} L_{pp} \sim
10^{31}\; \CMS.
\end{equation}
This is too small for study of any good physics.

  However, one can think about a special source of proton with several
stages of  electron cooling. If parameters of the proton beam are the
same as those of the muon beam, then the luminosity at the 100 TeV
$\mu p$ collider $L_{\mu p} = \LMM \sim 10^{36} \; \CMS$.
 That is not easy to achieve, but such possibility is not excluded. 

One of problems at such colliders is hadronic background. At $L_{\mu
p} = 10^{36}$ and $\nu = 10^4$ the number of background $\gamma p$
reactions is about 5000/crossing. One can decrease backgrounds
by increasing the  collision rate (up to a factor of 5--10). It is not
excluded that even with such backgrounds one can extract interesting
physics. This option certainly makes sense, if a very high energy $\mu\mu$
collider is to be built. Its feasibility and potential problems should be
studied in more detail.

\end{document}